\documentclass[12pt]{article}
\usepackage{color}
\usepackage{graphicx}
\usepackage{latexsym}
\usepackage{amsmath}
\usepackage{amsfonts} 
\usepackage{amssymb}
\usepackage{indentfirst} 
\numberwithin{equation}{section}
\newcommand{\be}{\begin{equation}}
\newcommand{\ee}{\end{equation}}

\newcommand{\mE}{{\mathcal E}}
\newcommand{\mG}{{\mathcal G}}
\newcommand{\mF}{{\mathcal F}}
\newcommand{\mL}{{\mathcal L}}

\newcommand{\mA}{{\mathcal A}}

\title{From polarized gravitational waves to   analytically solvable  electromagnetic beams\\
}
\author{K. Andrzejewski\footnote{Corresponding author, e-mail: k-andrzejewski@uni.lodz.pl},
 \quad S. Prencel
\vspace*{0.8cm}
\\
\small  Department of  Computer Science,  Faculty of Physics    and Applied Informatics, \\ \small
 University of  Lodz, Pomorska 149/153,
90-236, Lodz, Poland\\
}
\date{}
\begin{document}
\maketitle 
\begin{abstract}  Using  the correspondence between  solutions of  gravitational and gauge theories   (the    so-called  classical    double copy  conjecture)   
some   electromagnetic fields with  vortices are  constructed, for which   the Lorentz force  equations are  analytically solvable.  The starting   point is    a certain  class of  plane gravitational waves  exhibiting  the conformal symmetry.   The notion of the Niederer transformation, crucial for the  solvability,   is analysed in the case of the   Lorentz force equation  on the  curved spacetimes
as well as its  derivation by means of   integrals of motion (associated with  conformal generators preserving these vortices) is presented.
Furthermore, some  models  discussed recently  in the context  of  the intense laser beams are constructed from their  gravitational counterparts, with the special emphasis put on the focusing property, and  new  solvable examples are presented.   
\end{abstract}
\section{Introduction}
\label{s1}
Gravitational waves  have been intensively studied  since the invention of general relativity. Recently they have gained  a new interest, both due to their  direct observations 
 \cite{b1h,b1i} as well as  because of  new  theoretical ideas  like the memory effect and  soft graviton theorems (\cite{b3d}-\cite{b4bbb} and  references therein).  The circularly  polarized ones seems to be particularly  interesting  as  they  may  arise as  the effect  of  coalescing black holes,  neutron star merger or can be observed from the  astrometric data \cite{b1i,b4c}.  The linearly polarized waves are, in turn, distinguished  by their  relative  simplicity. Far from the  source in the  neighbourhood  of the detector one can approximate gravitational  waves by  the exact plane ones  (assuming that the back reaction of the detector  is negligible).  Finally,  the  so-called  impulsive gravitational  waves seem to be of some importance   \cite{b5e}-\cite{b7}.
\par 
{
On the other hand, it turns out  that  in the case of  the  exact gravitational waves  mentioned above   there are some three special classes; they are  defined by  the maximal (in  the non-flat case), seven-dimensional, conformal symmetry \cite{b6a,b6aa,b6b}; they are the only  ones exhibiting the maximal conformal symmetry among all nontrivial vacuum solutions to the Einstein equations.  One of these classes   consists of a  metric family  describing   linearly or circularly polarized gravitational pulses. This  exceptional family  admits the  proper conformal transformations; moreover,  it  can be used to model  impulsive gravitational waves with the  Dirac delta profile. However, the most interesting property of this family is that the geodesic equations can be explicitly solved. In consequence,   some  gravitational phenomena (such as  singularities,    focusing,   classical cross section and the velocity memory effect)   can be   analysed analytically, also in the Dirac delta limit \cite{b7,b77}. Such a situation is   strictly   related    to the existence of the so-called  Niederer transformation \cite{b77,b7a,b7b} (see also  Sec. \ref{s2} for some details).}
\par In this work, based on the mentioned above polarized  gravitational pulses,   we study the interaction of the  electromagnetic field with a charged particle.  The bridge between the  gravitational   waves and electromagnetic fields  is provided by   the idea of the classical  double copy  (a part of the colour-kinematic duality, see  \cite{b8a}-\cite{b8f} and references  therein). At the quantum level (analysed as the tree and few loops levels) this conjecture concerns the problem of  how scattering amplitudes in gravity can be obtained from those in the  gauge theory by replacing colour structure with kinematical one. At the classical level,  it consists in   the mapping of the solutions of the Einstein equations into the  solutions of  the Yang-Mills equations; for example a  Kerr-Schild  type relation  of the  double copy \cite{b8a} where the  linear structure of the Einstein equations for the Kerr-Schild metrics   corresponds to  Abelian  gauge  fields. 
{ 
One of the main examples of such a correspondence  is provided by the plane gravitational waves; they correspond to the electromagnetic potentials which yield certain, non-plane and vacuum, electromagnetic fields. 
 As we indicated above there is an exceptional family of the  plane gravitational waves for which the geodesic equations can be analytically analysed. In view of the double copy conjecture the following question arises; does the same situation hold for the motion of a charged particle in the electromagnetic fields corresponding to this class? The positive answer to this question is obtained in Sec.  \ref{s3}. In consequence, by using of the idea of double copy we obtain electromagnetic  fields  for which (as for its gravitational counterparts) the dynamics is explicitly integrable.
 Moreover, the  fields obtained    exhibit vortices and generalize  the ones described  in Ref. \cite{b9} (which  seem to be of some  importance for singular optics and trapping problems, i.e.   the confinement mechanism of particles by electromagnetic or gravitational fields, see \cite{b9a, b9c} for the electromagnetic case and  \cite{b18,b18b,b18c} for its gravitational counterpart). It is also worth to notice that for  the  electromagnetic fields under consideration   the Dirac delta limit of the profile can be easily performed.    
 }
 \par
 Since the   solvability  discussed above    is  strictly related to  the notion of the Niederer transformation   we  discuss, in Sec. \ref{s4},     a geometric  extension of the Niederer  map to  the case of the plane gravitational spacetimes endowed with  some, crossed, electromagnetic  fields.  As a consequence we obtain some examples of  Lorentz force equations in suitable gravitational backgrounds which are  solvable in the transverse directions.
\par 
In Sec.  \ref{s5}  we slightly modify  the Kerr-Schild  ansatz to produce non-null electromagnetic fields. Such  electromagnetic fields  have been discussed in Refs. \cite{b10a,b10b};  they  capture   some essential features of the transverse magnetic beam of  laser light  near the beam axis.   Next, we apply this procedure  to  the linearly polarized  gravitational metrics    and obtain  an explicitly solvable  model  described in Ref. \cite{b10a}.  
Moreover,  relying on a   member of  the family of circularly polarized gravitational pulses we  construct  a transversally solvable electromagnetic background. However, in contrast to the previous one   it has  zeros; this fact  essentially  modifies the focusing conditions  important for intense  or ultra-short laser pulses  (when the paraxial approximation may be no longer valid, see  e.g.   \cite{b11a,b11b}).  In the  general case, we give some criteria  and  applied to the  cases under consideration. 
\par 
Moreover,  in Sec. 6 the role and meaning  of   integrals of motion  are discussed with the special emphasis on new ones associated with  conformal generators  preserving vortices under discussion;  it is shown that these integrals  lead to the  Ermakov-Lewis  invariants \cite{a0a,a0b,a0c} and consequently to  the Niederer transformation.
The  results outlined  above suggest deeper connection (related to the  conformal symmetry and integrability) between  gravitational and gauge theories  rooted in the double copy approach.        Finally, in Sec. \ref{s7} we give a summary with an outlook for further studies.
\section{Plane gravitational  waves and conformal symmetry}
\label{s2}
In this section we recall  some facts  crucial for our further studies. They concern the role of the conformal symmetry in the  problem of analytical solutions of the geodesic equations.     Namely,
let us consider a  subclass of the   pp-waves,  the so-called  generalized  plane gravitational   waves,    of the following  form\footnote{We use the following conventions: signature $(-,+,+,+)$ and $u\equiv (x^3-x^0) /\sqrt{2}$,  bold indices refer to 2-dimensional  vectors, the Einstein  convention is assumed.}
\be
\label{e000}
g={\bf x}\cdot H(u){\bf x}du^2+2dudv+d{\bf x}\cdot d{\bf x},
\ee
where $H$ is assumed, without loss of generality, to be   a symmetric matrix.  In general,  they are   solutions to Einstein's field equations in which the only source of  gravity is some kind of  radiation i.e.   the source is a null fluid. The weak energy condition implies  $\textrm {tr}(H)\leq 0$ and the  scalar curvature vanishes. If $\textrm {tr}(H)= 0$ then  $g$   satisfies  the vacuum Einstein equations  and,  consequently, describe  a  plane gravitational   wave (exact gravitational wave). 
The geodesic equations for  the metric (\ref{e000})  reduce to the following ones 
 \begin{align}
\label{e6a}
\overset{..}{{\bf x}}&=H {\bf x},\\
\label{e6b} \overset {..}{v}&=-\frac12 {\bf x}\cdot  \dot{H}{\bf x}-2 {\bf x}\cdot H	\dot {\bf  x},
\end{align}
where  dot refers to derivative with respect to $u$.  Furthermore, eq. (\ref{e6b}) can be  directly integrated  yielding
\be
\label{e7}
v(u)= -\frac  12 {\bf x}\cdot \dot {\bf x}+C_1u+C_2,
\ee
where ${\bf x}$ is a   solution to  (\ref{e6a}).  
Thus for  the (generalized)  plane  gravitational waves the solution to   geodesic  equations is obtained by solving  the set of eqs. (\ref{e6a}). Moreover,   the latter  are particularly  interesting in the case of the exact   gravitational waves since they  coincide with  the deviation equations and enter  the   transformation rules to the  so-called  Baldwin-Jeffery-Rosen  coordinates \cite{b2c,b2d}.  Although this  set of equations cannot,  in general,   be explicitly solved,  there are some special cases when the solution   is accessible (see  e.g., the classical paper \cite{b20}).  Another, more geometric  approach to this problem  is  related to the symmetry of the metric; the most   interesting cases are the ones exhibiting the maximal symmetry. 
Let us start with the  isometry  groups. It is well known that the generic dimension  of the isometry  group of plane gravitational waves is five. There  exist two exceptional  families admitting the  six-dimensional isometry groups. The first one  is defined by the matrix $H$  of the form 
\be
\label{ee3}
H^{(0)}(u) =\left(
\begin{array}{cc}
\cos(\kappa u) &\sin(\kappa u)\\
\sin(\kappa u) & -\cos(\kappa u)
\end{array}
\right);
\ee
it  admits  the explicit solutions to the geodesic equations (cf.   \cite{b14b,b14a} and references therein). The second  family  describes  geodesically incomplete manifolds  defined by $H\sim u^{-2}$;    it is intensively  used    in the context of  the Penrose limit \cite{b6f,b6ff}.
\par 
The situation becomes  more interesting if we take into account the conformal symmetry. Let us recall that  the  maximal dimension of the conformal group of the, non-conformally flat,  metric    is   seven  \cite{b6g}.   Note that the metric  (\ref{e000})  admits a homothetic vector field. Therefore,  the two, mentioned above, families   exhibit the seven-dimensional conformal  symmetry  (six isometries and homothety). It appears that  there  exists only one subclass of the plane gravitational waves  with the  seven-dimensional  conformal group consisting  of five-dimensional isometry, homothety and non-homothetic  conformal  transformations (cf.  \cite{b6a,b6aa,b6b} and \cite{b6c,b6d,b6e}).   This special subclass consists of two families describing  linearly and circularly polarized plane  gravitational waves. In the following,  we concentrate on them  (actually, on the geodesically complete  cases as the most interesting ones). The first family, linearly polarized,  is defined  by the     metric $g^{(1)}$ with     the profile 
\be 
\label{e8}
H^{(1)}(u)=\frac{a}{(u^2+\epsilon^2)^2} 
\left(
\begin{array}{cc}
1&0\\
0&- 1
\end{array}
\right),
\ee
where $ \epsilon>0$ and $a$ is an arbitrary number (excluding the trivial  Minkowski case and changing $x^1$ and $  x^2$  one can assume $a>0$). Moreover, let us note that  taking $a\sim \epsilon^3$  one obtains the impulsive gravitational wave with  the Dirac delta profile (as $\epsilon$ tends to zero).
\par 
The second family $g^{(2)}$  is an example of  the   circularly polarized  plane gravitational  waves. It is defined by the following profile 
\be 
\label{e9}
H^{(2)}(u)=\frac{a}{(u^2+\epsilon^2)^2}
\left(
\begin{array}{cc}
\cos(\phi(u)) &\sin(\phi(u))\\
\sin(\phi(u)) & -\cos(\phi(u))
\end{array}
\right),
\ee
where 
\be
\label{e9a}
\phi(u)=\frac{2\gamma}{\epsilon}\tan^{-1}(u/ \epsilon),
\ee
 $\epsilon, \gamma>0$  and  $a$ can be chosen as above  (for $\gamma=0$, eq. (\ref{e9})  reduces to the previous case; however, for  physical and mathematical reasons we shall consider linear and circular polarizations  separately).
\par 
Some properties  of  the gravitational waves  $g^{(1,2)}$ were  discussed in \cite{b7,b77}. Among others it was noticed  that   the geodesic equations  can be explicitly solved; furthermore, it was shown that this  fact can be simply explained in terms of   the so-called   Niederer transformation (see also Sec. \ref{s4}). In the case of $g^{(1)}$   the transversal  part of the  geodesics  reads 
\be
\label{e10}
 x^i(u)=C_1^i \sqrt{u^2+\epsilon^2}\sin(\Lambda_i\tan^{-1}( {u}/{\epsilon})+C_2^i),
\ee
where 
\be
\label{e11}
\Lambda _i=\sqrt{1+(-1)^i\frac{a}{\epsilon^2}} \qquad  i=1,2.
\ee
Moreover,  the initial conditions 
\be
\label{e21b}
\dot{\bf  x}(-\infty)=0,\qquad  {\bf x}(-\infty) ={\bf x}_{in},
\ee
lead to the  observation  that   only  the   second component $x^2(u)$ exhibits focusing.
In the case of  $g^{(2)}$,  the solutions are   of the form 
\be
\label{e2}
{\bf x}(u)= \frac{\epsilon R(\tilde u){\bf y}(\tilde u) }{\cos(\tilde u)}, \quad  \tilde u=\tan^{-1}(u/\epsilon),
\ee
where 
\be 
\label{e2b}
R(\tilde u)=
\left(
\begin{array}{cc}
\cos(\omega \tilde u)&-\sin(\omega \tilde u)\\
\sin(\omega \tilde u)&\cos(\omega \tilde u)\end{array}
\right)  \qquad \omega=\frac{\gamma}{\epsilon},
\ee
and $y$'s are solutions to the following set of differential equations  with constant coefficients
 \be
 \label{e17}
 \begin{split}
(y^{2}) ''  +2\omega (y^1)'+\Omega_- y^2&=0,\\
(y^{1})''   -2\omega (y^2)' +\Omega_+ y^1&=0,
 \end{split}
 \ee
 with
 \be
 \label{e18}
 \Omega_\pm=1-\omega^2\mp \Omega, \qquad \Omega=\frac{a}{\epsilon^2};
 \ee   
 here   primes  refer to the derivatives with respect to $\tilde u$.  
Although the   general solution of eqs. (\ref{e17}) is  well  known, in our case it is a  linear  combinations of  trigonometric functions only or both hyperbolic (or linear) and trigonometric ones depending on the values of parameters appearing in (\ref{e17}),  the    form of  the coefficients can be  quite  complicated (for $\gamma=\epsilon$  they  are presented in \cite{b77}). In the general case  the  transformation to the normal coordinates seems more useful (see  also \cite{b15a,b15b}). 
 The above   results  form a  setup for  our further considerations. 
\par 
\section{Exactly solvable electromagnetic vortices}
\label{s3}
{In the previous section we pointed out   that for the plane gravitational pulses exhibiting the  maximal conformal symmetry the geodesic equations can  be analytically  solved. Here, we showed that a similar situation holds for the electromagnetic fields constructed by means of the double copy conjecture, i.e.  the existence of explicit solutions to the geodesic equations coincide with the existence of the explicit solutions to the Lorentz force equations for the corresponding electromagnetic backgrounds.  
}
To this end let us consider, in the Minkowski  spacetime and light-cone coordinates,   the following electromagnetic one-form  
\be
\label{e23}
\mA= -{\bf x}\cdot A(u){\bf x}du .
\ee
Such a potential yields  the  (crossed)   electromagnetic  fields 
\be
\label{e25}
\vec E= (f_1,f_2,0), \quad \vec B=(-f_2,f_1,0),
\ee  
where 
\be
\label{e26}
{\bf f}=(f_1,f_2)=(\sqrt2 A_{1i}x^i,  \sqrt{2}A_{2i}x^i).
\ee
The fields   $\vec E $ and $\vec B$  satisfy Maxwell's equations with the following null current 
\be
\label{e28}
j^\mu=\sqrt 2 {\textrm {tr}}(A)(1,0,0,1), \quad j^\mu j_\mu=0,  
\ee
(cf.  \cite{b17a,b17b}). 
\par 
 Furthermore,  the following conditions 
\be
\label{e27}
\vec E^2-\vec B^2=0, \quad \vec E\cdot \vec B=0,
\ee
hold and imply  that the electromagnetic field  is the pure radiation (the energy density and the Poynting vector form a null four-vector).  
When ${\textrm{ tr}}(A)\neq 0$, e.g. $A$ is proportional to the  identity,  then  $j^\mu$  can be interpreted as  a pulse of charges of one signs  moving with the 
speed of light   along $z$-axis, especially when a suitable  regularization  is performed  ($j^\mu$ is zero outside  a transversal  region); unfortunately,  even then the total  energy of fields is infinite. The situation changes  if  a dipole model of the   particle is taken  under consideration, for more details see the original paper \cite{x1}. 
\par 
The case   $\textrm{tr}(A)=0$ corresponds   to the electromagnetic field satisfying the  vacuum Maxwell equations; however, 
 such fields are not, in general,  electromagnetic plane waves.  Moreover, for arbitrary  vacuum  solutions the    conditions  (\ref{e27})   (equivalently 
the vanishing of the square of  the Riemann-Silberstein vector)   can be used to describe   the  vortex of the electromagnetic field  \cite{b16a}. However,  in the null fluid case, i.e.  when  (\ref{e27}) vanishes identically,    the  notion of the  vortex can be  simplified;   it is then    related to  the condition that all components of the electromagnetic field vanish  (see  \cite{b16b}). In this approach the   electromagnetic field  (\ref{e25}) carries a straight vortex line along the $z$-axis. 
{This is interesting due to the  fact that the  electromagnetic vortices (vortex lines)   gained some attention not only in the context  of singular optics  but also  in the confinement mechanisms of particles by electromagnetic  fields  or even knot  theory, see   \cite{b9,b9a, b9c} and references therein. 
}
\par 
 Now let us note that for the potential (\ref{e23})  the  Lorentz force equations
\be
\label{e29}
m\frac{d^2x^\mu}{d\tau^2}=e{F^\mu} _{\nu}\frac{dx^\nu}{d\tau},
\ee
 give
\begin{align}
\label{e30}
m\frac{d^2{\bf x}}{d\tau^2}&=e{\bf f}\frac{d}{d\tau}(x^0-x^3),\\
\label{e31}
\frac{1}{\sqrt{2}}(\frac{dx^0}{d\tau}-\frac{dx^3}{d\tau})&={-}\frac{d u}{d\tau}= \frac{-p_v}{m}>0;
\end{align}
($p_v<0$ is the light-cone momentum  of the particle)   and consequently  eqs. (\ref{e29})  can be expressed in terms of the  $u$ coordinate as follows 
 \begin{align}
 \label{e32a}  \overset{..} {\bf x}&=\frac{2e}{-p_v}A{\bf x},\\
\label{e32b}\overset{..}{ v}&=\frac{2e}{p_v}{\bf x}\cdot A\dot {\bf x},
\end{align}
(see also \cite{b17a,b17b}). Thus, up to the constant $\frac{2e}{-p_v}$,  the transverse part of the equation  of motion  (\ref{e32a}) has the same form  (in contrast to the longitudinal direction, see  the discussion below) as for the plane gravitational waves (cf.   eqs.   (\ref{e6a})) . 
\par
The above properties of the vacuum electromagnetic field given by  the  potential (\ref{e23}) can be considered  as  a manifestation of  the  idea of double copy which at the quantum  level  concerns the problem of how scattering amplitudes in gravity can be obtained from those in the gauge theory by replacing colour structure with kinematical one; for more details we refer to some surveys \cite{b8d1,b8d2} because  our further considerations will focus on its classical counterparts which have been  proposed (see e.g.    \cite{b8a}-\cite{b8d})  to better understand this  conjecture  (for example whether  the copy is a genuinely non-perturbative property of both theories). In this   approach we  look directly at solutions of the classical field  equations in gauge and gravity theories and  match these up according to a double copy prescription. Such a matching has been   observed for the metrics in the  so-called the Kerr-Schild coordinates \cite{b8a}; more precisely,  for   the metrics  which  can be written in the following form  
\be
\label{x00}
g_{\mu\nu}=\bar g_{\mu\nu}+\lambda k_\mu k_\nu,
\ee
where $\lambda$ is a scalar function and $k^\mu$ is s a null vector with respect to both metrics.
It turns out that under assumption $k^\mu \bar D_\mu k_\nu$=0  the Ricci tensor has the remarkable property that it is linear in $\lambda$.  This property suggests that the corresponding gauge field should take the most simple form (even  an Abelian one). For the Minkowski background $\bar g=\eta$   the  following (in general non-Abelian ansatz) was proposed 
\be
\label{x0}
A^a_{\mu} =c^a \lambda k_\mu,
\ee 
where $c^a$ is an arbitrary constant colour vector. In this case  under some additional assumptions concerning $\lambda$ and $k_\mu$  (e.g. stationary)  vacuum solutions are invariant  under this correspondence. 
Although it  is not known  how to  fully extend the  above case    several attempts have been   made (including   AdS background, Taub-Nut or Kundt spacetimes, see \cite{b8b,b8c,b8e,b8f}).  One of the most natural extensions is based on  the metric of the following form 
\be
\label{x0a}
g_{\mu\nu}=\bar g_{\mu\nu}+\lambda_1 k_\mu k_\nu+ \lambda _2 l_\mu l_\nu,
\ee
where $l^\mu$   is also a null vector   and orthogonal,  w.r.t. $g$ and $\bar g$,  to $k^\mu$. For such metric the corresponding gauge potential  (in  $\bar g$ spacetime) reads
\be
A^a_{\mu} =c^a( \lambda_1 k_\mu +\lambda_2 l_\mu ).
\ee
One of the  main examples of  the correspondence above  outlined    is provided by  the plane gravitational waves and  non-plane electromagnetic fields. Namely, taking $\bar g=\eta$ and  the vector $k^\mu=(1,0,0,1)/\sqrt 2$,   the generalized plane wave (\ref{e000})   takes the form (\ref{x00})  with $\lambda=\lambda(u,\vec x)={\bf x}\cdot H(u){\bf x}$.  According   to the classical double copy  conjecture  the corresponding  electromagnetic potential  (\ref{x0})  coincides with  (\ref{e23})    if  we identify   the matrices  $A=H$ (in the most simple, Abelian, case).  In Sec. \ref{s4}  we  use this correspondence  taking a  plane wave spacetime  (\ref{e000}) as  a fixed background $\bar g$;  then the additional term in (\ref{x00})  can be interpreted as the potential in curved spacetime $\bar g$ (cf.  (\ref{x0a}) and   see also a more detailed discussion in Refs.  \cite{b8b,b8c}). 
\par   
The above outlined reasoning has been  used,  in Ref.  \cite{b18}, to study    an electromagnetic  vortex proposed in  \cite{b9} which  can act as a beam guide for charged particles; moreover, it  is analytically solvable (see also \cite{b17a})  as well as provides  an approximation to  more realistic beams \cite{b9b}.  Such a vortex   corresponds to     the plane gravitational wave,  defined by the profile (\ref{ee3}), which  is also analytically solvable.  Guided by  the above    idea of classical double copy  in what follows  we show that the  vortex mentioned   possesses   solvable  extensions based on the gravitational waves related to  the proper  conformal symmetry discussed in the previous section. 
\par 
To this end  let us consider the electromagnetic potentials   related to the   plane gravitational waves       $H^{(1,2)}$. Then, by virtue of    (\ref{e8}), (\ref{e9}) and    (\ref{e23})-(\ref{e26}), one obtains the following    electromagnetic  fields      in   the Minkowski  spacetime
\be
\label{e33}
\vec E^{(1)}(x)= \frac{\sqrt{ 2} a}{\left(u^2+\epsilon^2\right)^2}\left(x^1,-x^2,0\right),\quad \vec B^{(1)}(x)= \frac{\sqrt{ 2} a}{\left(u^2+\epsilon^2\right)^2} \left(x^2,x^1,0\right);
\ee
\be
\label{e34}
\begin{split}
\vec E^{(2)}(x)&= \frac{\sqrt{ 2} a}{\left(u^2+\epsilon^2\right)^2}\left(x^1\cos(\phi(u))+x^2\sin(\phi(u)),x^1\sin(\phi(u))-x^2\cos(\phi(u)),0\right),\\
\vec B^{(2)}(x)&= \frac{\sqrt{ 2} a}{\left(u^2+\epsilon^2\right)^2}\left(-x^1\sin(\phi(u))+x^2\cos(\phi(u)),x^1\cos(\phi(u))+x^2\sin(\phi(u)),0\right),
\end{split}
\ee
where $\phi$ is given by (\ref{e9a}) and  $u=(x^3-x^0)/\sqrt{2}$. As it  has been   indicated   in Sec. \ref{s2}   the geodesics equations   for  $g^{(1,2)}$  are analytically solvable; in consequence    one immediately (after replacing $a\mapsto \frac{2e}{-p_v}a$, see (\ref{e6a}) and (\ref{e32a}))  obtains the solution to the  transverse  Lorentz force  equations with $\vec E^{(1,2)}$ and $\vec B^{(1,2)}$ (cf. eqs. (\ref{e10}) and (\ref{e2})-(\ref{e18})).  Finally, taking an appropriate limit (cf. \cite{b7,b77})   one gets the case of electromagnetic field with the profile  proportional to  the  Dirac delta function $\delta(u)$.
\par Now,  let us   consider the longitudinal direction. In contrast to the gravitational case eq. (\ref{e32b})  can be  directly  integrated only once
\be
\label{e35}
\dot v=-\frac{1}{2}\dot {\bf x}\cdot \dot {\bf x}+D_1.
\ee
However, in what follows  we show that for the field given by  (\ref{e33})  and (\ref{e34})    it is also possible to find  explicitly   the $v$ coordinate. Indeed,  integrating (\ref{e35}) by substitution $\tilde u= \tan^{-1}(u/\epsilon)$  one obtains 
\be
\label{e36}
v(u)=D_1u+D_2-\frac{1}{2\epsilon}\int {{\bf x}'(\tilde u)}\cdot  {\bf x}'(\tilde u)\cos^2(\tilde u)d\tilde u.
\ee
Now, for the fields  $\vec E^{(1)},\vec B^{(1)}$ using (\ref{e10}) one gets,    after some computations,  the following final form of the longitudinal coordinate
\be
\label{e37}
\begin{split}
v(u)&= -\frac {\epsilon}{2} \sum_i(C_1^i)^2\Big(\frac 1 2(b_i^2-1)\tilde u+\sin^2(b_i\tilde u+C^i_2)\tan(\tilde u) +\frac{1+b_i^2}{4b_i}\sin(2(b_i\tilde u+C_2^i))\Big)\\
& +D_2+D_1u,
\end{split}
\ee
where $\tilde u= \tan^{-1}(u/\epsilon)$ and $b_i=\sqrt{1-(-1)^i\frac{2ea}{\epsilon^2p_v}}$.
\par 
The second case  is slightly more involved. 
First, we express  the   integral in (\ref{e36})  in terms of $y$'s satisfying   eqs.  (\ref{e17}) (with $\Omega=\frac{2ea}{-p_v\epsilon^2}$) and, subsequently, we extract a total time derivative term;  in consequence we arrive   at the following equalities 
\begin{align}
\label{e5}
&-\frac{1}{\epsilon^2}\int {{\bf x}'(\tilde u)}\cdot  {\bf x}'(\tilde u)\cos^2(\tilde u)d\tilde u=\nonumber \\
&\int \left[\left ((y^1)'-\omega y^2\right)^2+\left((y^2)'+\omega y^1\right)^2+2\tan(\tilde u)((y^1)'y^1+(y^2)'y^2)+\tan^2(\tilde u)((y^1)^2+(y^2)^2)\right]d\tilde u\nonumber\\
&=(y^2)'y^2+(y^1)'y^1+((y^1)^2+(y^2)^2)\tan(\tilde u)+\Omega\int [(y^2)^2-(y^1)^2)]d\tilde u.
\end{align}  
Let us note  that $y$'s are   combinations of the trigonometric or hyperbolic functions (cf.   (\ref{e17})); thus the last integral in (\ref{e5})   is an elementary one and can be explicitly computed. Finally, substituting $\tilde u=\tan^{-1}(u/\epsilon)$  in (\ref{e36})   one obtains the form of $v(u)$  which     gives the   analytical solvability of the Lorentz force  equations in the electromagnetic backgrounds  (\ref{e33}) and  (\ref{e34}) (extending  in this way  some  results    obtained in Ref. \cite{b9}  to a kind of electromagnetic pulses  exhibiting vortices).   
\section{Niederer's map  and  Lorentz's force equation}
\label{s4}
It turns out that \cite{b77} the existence of analytical solutions to the geodesic equations for the gravitational waves $g^{(1,2)}$   can be simply explained  in terms of  the  so-called  Niederer transformation \cite{b7a,b7b}. 
We shall show that the similar  situation holds  also when  the  electromagnetic  fields  (\ref{e33}) or (\ref{e34})  are   switched on. To this end 
{
 and to make the paper more self-contained let us recall  some facts concerning the Niederer map and its geometric interpretation}. The Niederer transformation\footnote{For    further  considerations  we adopt  
$u$-notation. There exists  a  hyperbolic  counterpart of  Niederer's transformation leading to the repulsive case.}
\be
\label{e12}
\begin{split}
u&=\epsilon\tan(\tilde u), \\
{\bf x}&=\frac{\epsilon \tilde {\bf x}}{\cos(\tilde u)},\\
\end{split}
\ee
 relates the free motion $\overset{..}{{\bf x}}=0$ (for our purpose we consider  the 2-dimensional case)  on the whole real axis ($-\infty<u<\infty$) to the half of period motion ($-\frac\pi 2<\tilde u<\frac \pi 2$) of the  attractive harmonic motion $\tilde{\bf x}''=-\tilde {\bf x}$;  as above   dot and prime  refer to the derivatives with respect to $u$ and $\tilde u$, respectively (this equivalence continues  also  to   hold at the quantum level  \cite{b7a}). 
 Of course,  the above  observation  have a local character;   however,  it reflects a similarity between  both systems  and brings  some useful information. 
 Various local quantities   can be directly related; this  concerns  even the global  ones (for instance, Feynman propagators) if sufficient care is exercised (see, e.g.,  \cite{b7a,b7c,b7d}).   In particular,   the  maximal symmetry groups of both systems   are isomorphic and one obtains the explicit relation between  symmetry generators as well as   solutions  of  the both systems \cite{b7e}. 
\par  
On the other hand eqs. (\ref{e6a})  describe, in fact,   a linear oscillator (in general)  with   time-dependent frequencies; however,   it turns out that   in some cases the Niederer  mapping  can be  also    applied to relate  them to  a harmonic (or a  more simple linear) oscillator  \cite{b77}.  Namely,  under the Niederer transformation eqs. (\ref{e6a})  are transformed into the following ones 
\be
\label{e00g}
\tilde {\bf x}''=\tilde H(\tilde u)\tilde {\bf x},
\ee 
where 
\be
\label{e00d}
\tilde H(\tilde u)=\frac{\epsilon^2H(\epsilon \tan(\tilde u))}{\cos^4(\tilde u)}-I.
\ee
In particular, if  the matrix  $H$   is of the form 
\be
\label{e00k}
 H(u)=\frac{a}{(\epsilon^2+u^2)^{2}}G(u),
\ee
where $G$ is a symmetric matrix, then,  by virtue  of eq. (\ref{e00d}),  the equations of motion   (\ref{e00g}) read
\be
\tilde {\bf x}''=\tilde H(\tilde u)\tilde {\bf x}=\left(\frac{a}{\epsilon^2} G(\epsilon \tan(\tilde u))-I\right) \tilde {\bf x}.
\ee
For example,     the (non-singular) time-dependent  linear  oscillator defined by a constant matrix $G$  is mapped under (\ref{e12})    to a part of   motion of the  harmonic oscillator.
In consequence   the Niederer transformation can be useful to solve  some  geodesic equations for plane  gravitational waves (or other problems  where time dependent linear oscillators occur). Moreover, such properties of the Niederer transformation have a reflection in a   geometric picture obtained by means of the Eisenhart-Duval lift  \cite{b7b,a3} (see also \cite{a4,a5} and references therein).  Namely, extending  the Niederer map by  adding the following  transformation rule 
\be
 \label{e12b}
 v=\epsilon\tilde v-\frac {\epsilon \tan(\tilde u) }{ 2}\tilde {\bf x}^2,
 \ee 
 one   gets the identity 
\be
\label{e00e}
g\equiv {\bf x}\cdot H(u){\bf x}du^2+2dudv+d{\bf x}^2=\frac{\epsilon^2}{\cos^2(\tilde u)}(\tilde {\bf x}\cdot \tilde H(\tilde u)\tilde {\bf x}d\tilde u^2+2d\tilde ud\tilde v+d\tilde {\bf x}^2)\equiv \frac{\epsilon^2}{\cos^2(\tilde u)}\tilde g,
\ee
where $H$ and $\tilde H$  are connected by eq. (\ref{e00d}). 
For the  particular case, $H=0$, (\ref{e00e})  reduces to the well-known   relation   between the  Bargmann spacetimes  corresponding to the free $(u,{\bf x},v)$ and  the half-oscillatory period  $(\tilde u,\tilde {\bf x},\tilde v)$ motions in the    Eisenhart-Duval lift language  \cite{b7b} (see also \cite{a6,a6b}). 
 \par 
 In view of  eq. (\ref{e00e}), by means of the Niederer transformation  one can associate with the metric $g$   the new one   $\tilde g$,  conformally related  to $g$, belonging to the same class (generalized plane gravitational waves). Of course,   at most only  one of the  metrics $g$ and $\tilde g$  describes  the vacuum  solution; moreover, the   geodesic  equations  for $g$ and $\tilde g$ are not  equivalent  (except  those for null geodesics). However, from the reasoning underlying the Niederer transformation it follows that the transversal geodesic  equations (\ref{e6a}) for  $g$  are mapped into the transversal geodesic equations (\ref{e00g})  for  the metric $\tilde g$,  which may be more tractable than  those for $g$.    The non-equivalence of the  geodesic equations for  $g$ and $\tilde g$  is reduced  to  the non-equivalence  of equations  determining  $v$ and $\tilde v$ (however,  the latter can be easily  solved,  cf. eq. (\ref{e7})).    Such a situation  holds, for instance,   for the conformally distinguished metric  families:   $g^{(1,2)}$ and the one  defined by  $G\sim I$ (see \cite{b7,b77}). 
\par
{Now, we extend the  geometric picture of the Niederer transformation   outlined  above to the case of both gravitational and electromagnetic  backgrounds; as a consequence we obtain some examples  of the Lorentz force equations in suitable gravitational backgrounds which are  solvable in the transverse directions. }
Namely, let  us analyse the action of  Niederer's transformation   on the spacetime described  by the metric (\ref{e000})   (in particular, for  $H=0$  -- the Minkowski one)  endowed  with   the  electromagnetic field   given by the potential (\ref{e23}).  First let  us note that  the condition $\textrm{tr}(A)=0$  implies that, as in the Minkowski case,     the  electromagnetic field under consideration satisfies the  vacuum Maxwell equations on the spacetime defined by   a plane gravitational wave  (this fact is valid even in the case of  an arbitrary  pp-wave spacetime). Furthermore,  in this case  the equations of motion  of a  charged test particle  
\be
\frac{d^2x^\alpha}{d\tau ^2}+\Gamma^\alpha_{\nu\mu}\frac{dx^\nu}{d\tau}\frac{d x^\mu}{d\tau}=\frac q m g^{\alpha \nu}F_{\nu\mu}\frac{d x^\mu}{d\tau},
\ee
reduce to the  following ones
\begin{align}
\label{ee1} \overset{..}{{\bf x}}&=(H-\frac{2e}{p_v}A){\bf x}\\
\overset {..}{v}&=-\frac12 {\bf x}\cdot  \dot{H}{\bf x}-2{\bf x}\cdot (  H	-\frac{e}{p_v}  A)\dot  {\bf x}
\end{align}
(for the Minkowski spacetime they  coincide with the eqs. (\ref{e32a}) and (\ref{e32b})).
 \par
From the above we see that the transversal part of the  Lorentz  force equation is still decoupled and can be considered separately. Thus we can try to extend the geometric  interpretation  and applications of Niederer's transformation to   electromagnetic fields in  curved spacetimes. 
To this end let us note that   under the Niederer transformation   the  transverse set  of equations (\ref{ee1})   is mapped to  following ones
\be
\label{ee2}
\tilde {\bf x}''=\left(\tilde H(\tilde u)+\frac{2e}{p_v}\tilde A(\tilde u)\right)\tilde {\bf x},
\ee
where $\tilde H$ is given by (\ref{e00d}) while  
\be
\label{ee4}
\tilde A(\tilde u)=\frac{\epsilon^2}{\cos^4(\tilde u)} A(\epsilon \tan (\tilde u)) .
\ee
The solutions  to eqs. (\ref{ee2})    corresponding   the purely  gravitational case, i.e.  $A=0$  and $g=g^{(1,2)}$,  have been  discussed in   \cite{b7,b77} (see  Sec. \ref{s2}), while     for the Minkowski spacetime endowed with  the  electromagnetic field   given by (\ref{e33})  and (\ref{e34}) in Sec. \ref{s3}.  However, by  considering  these   gravitational and  electromagnetic backgrounds  altogether  (i.e. 
 $\vec E^{(i)},\vec B^{(i)}$ on the spacetime $g^{(i)}$,  for $i=1,2$ respectively) one   finds  that  in these cases the transverse  Lorentz force   equations, given by   (\ref{ee2}),   are also analytically  solvable.  
\par 
Next,   by virtue of (\ref{e00e})  the Niederer map    transforms    the metric $g$  conformally  into the metric  $\tilde g$  of the same   type, i.e. we do not leave out the  class of the  generalized plane gravitational waves;   for  the electromagnetic potential (\ref{e23})  one obtains 
\be
\mA= -{\bf x}\cdot A(u){\bf x}du=\epsilon \tilde {\bf x} \cdot \tilde A(\tilde u)\tilde  {\bf x} d\tilde u,
\ee 
where $\tilde A$ is defined by (\ref{ee4}).  Thus  the  potential $\tilde A$, in the new coordinates, is  also of the same type   as $A$ and,  consequently, yields  also  a crossed electromagnetic field.    Moreover, the pair $(\tilde g,\tilde A)$  is the one     for which the transverse part of  the Lorentz force  equations is given by    eq.  (\ref{ee2}) (let us note that  since $\tilde g$ is conformally related to $g$  the electromagnetic field arising from the potential $\tilde A$ is a vacuum solution to the Maxwell equation with respect to $\tilde g$; in contrast to the metric   case  where the vacuum solution  $g$ is mapped to the  non-vacuum null-fluid  solution $\tilde g$). 
 In this way we extend the notion  of the Niederer transformation,  originally established for the non-relativistic dynamical systems,  to the one including   electromagnetic fields on curved spacetimes;  namely    with the  pair  $(g,A)$    we associate the new  one $(\tilde g,\tilde A)$  such that the transverse part of the Lorentz force  equations  for $(g,A)$   transform, by means of the Niederer map,  into  the one for $(\tilde g,\tilde A)$ (moreover  $g,\tilde g$   and $A,\tilde A$ belong to the same classes: the generalized plane  gravitational waves and vacuum crossed electromagnetic fields, respectively).
\section{Light-matter interaction}
\label{s5}
{In this section we touch upon some problems of the interaction between  optical beams (such as  lasers)   and charged particles. To this end  let us recall that in  the  standard approach  the optical beams (including lasers beams or pulses) are described  by the plane electromagnetic waves. Thus the dynamics of a charged particle in the presence of laser pulses reduces, at the classical level, to the solution of the Lorentz force equations and, at the quantum level, to the Dirac (Klein-Gordon) equation or, in general, QED process, in these fields. However, the real laser beam is localized (a finite beam width) and it has high amplitude near propagating direction and less far away (transverse spatial variations); moreover, the beam width increases along the optical axis, thus the wave fronts show spherical nature. In consequence, a non plane wave description of such beams is necessary. The most popular solutions of this problem are based on the paraxial approximation of the wave equation (small angle between the wave vector and the optical axis). Under this assumption various solutions (such as the  Gauss beam, Hermite-Gaussian, Laguerre-Gaussian modes and others) were extensively studied. However, for  pulses with very short duration or ultra intense lasers (higher focusing increases the diffraction angle and intensity) the paraxial approximation can be no longer valid, see e.g.  \cite{b11a,b11b}  and references therein. Thus it would be desirable to go beyond the paraxial approximation. Various approaches to the solution of this problem were proposed. In Ref. \cite{b10a} the authors construct auxiliary electromagnetic backgrounds which captures some essential features of transverse magnetic beams  near the beam axis. Next, they show that the dynamics of a charged particle in such models can be explicitly discussed (without approximation); in this way they obtain some insight into predictions and analyses for experiments with intense or ultra-short lasers.  
More precisely, the following electromagnetic fields, see    \cite{b10a}, }
\be
\label{e38}
\vec E(x)=\mE(u)(x^1,x^2,\sqrt 2 u), \quad \vec B(x)=\mE(u)(-x^2,x^1,0),
\ee
were considered, where the function  $\mE(u)$ is  picked, at least,  at $u=0$.  Such  fields  capture some essential features of the transverse magnetic beams near the beam axis, such as  the polarization structure, the local rise of the transverse fields, and the suppression of the longitudinal field as well as satisfies $\vec E\cdot\vec B=0$ while, in contrast to (\ref{e27}),  $\vec E^2-\vec B^2>0$ (see also \cite{b10a}).   
\par
{In what follows, first, we shall show that }such fields can emerge  from the  gravitational metrics if we slightly modify  the   correspondence between gravitational, given by  (\ref{e000}), and electromagnetic fields. 
As before   we put  $A=H$  in the  one-form (\ref{e23}); however,  this time  we add a new term with an arbitrary function  $\mF$   
\be
\label{e39}
\hat \mA = -{\bf x}\cdot H(u){\bf x}du  +\mF(u)dv.
\ee
Then one gets
\be
\label{e40}
\vec E= (f_1,f_2,\dot\mF), \quad \vec B=(-f_2,f_1,0),
\ee  
\be
\label{e41}
j^\mu=\sqrt 2( {\textrm{ tr}}(H)+\frac 12\overset{..}{ \mF})(1,0,0,1), \quad j^\mu j_\mu=0,  
\ee
where $f_1$ and $f_2$ are given by (\ref{e26}). 
Now,  repeating the previous considerations (i.e. using  the Lorentz equations (\ref{e29}) and next integrating)  one finds the following relation 
\be
\label{e42}
m\frac{du}{d\tau}={-e\mF(u)-D },
\ee
 where  $D$ is a constant of  integration  (see also  (\ref{r4})). Then  the Lorentz    equations in terms of the $u$  coordinate read (cf.  \cite{b17a}) 
 \begin{align}
 \label{e43a} \overset{..}{ {\bf x} }(e\mF+D)+e\dot \mF\dot {\bf x}&={2e}H{\bf x},\\
 \label{e43b} \overset{..}{ v}(e\mF+D)+2e\dot \mF\dot v&=-{2e}{\bf x}\cdot H \dot{\bf x}.
\end{align}
\par
Let us consider the diagonal profile $H_{11}=H_{22}$,  $H_{12}=H_{21}=0$  and  the function 
\be
\label{e44}
\mF(u)=2\int_{-\infty}^{ u}\bar   uH_{11}(\bar u)d\bar u
\ee
(we assume here the   vanishing of the  gravitational profile at plus/minus null infinity in such a way that  the function $\mF$ is well defined). 
Consequently,  one gets  the electromagnetic field given by (\ref{e38})  with  
\be
\label{e44b}
\mE(u)=\sqrt{2}H_{11}(u).
\ee  
Of course, the    function $\mF$  can be chosen up to a (irrelevant)  constant, and our choice $\mF(-\infty)=0$, due to the asymptotic  vanishing of the gravitational  profile and (\ref{e42}),    gives  
 \be
D=-p_v,
\ee
where  $p_v<0$  (cf. (\ref{e31})). 
Finally, let us note  that   the  choice of  the  matrix  $H$,   $\textrm{tr}(H)\neq 0$,  corresponds to the, non-vacuum,      generalized   plane gravitational  waves (here, in general,   we do not assume  that  the weak energy condition holds)  and, consequently, gives  non-vacuum electromagnetic fields.  
\par 
The general solution of eqs. (\ref{e43a}) was obtained  in Ref. \cite{b10a} by means of some integrals of motion. Here we  apply slightly different method.  First,  let us recall that having a particular solution to the  second-order  homogenous linear differential equation,   the general solution can be expressed  in terms of some integrals containing this solution.    For a  diagonal  matrix $H$ and (\ref{e44})  the functions    $x^i_1(u)=u$  are   particular solutions  to the transverse  Lorentz force  equation (\ref{e43a}).  Thus using  the above property of ordinary differential equations   we obtain that the general solution is a linear combination of $x^i_1$ and the 
second solution of the form    
\be
\label{e45}
x^i_2(u)=u\int \frac{du}{u^2(e\mF(u)-p_v)};
\ee
(in the case of the longitudinal   direction  eq. (\ref{e43b})  can be  integrated once \cite{b10a}). 
\par
\begin{figure}[!ht]
\begin{center}
\includegraphics[width=0.45\columnwidth]{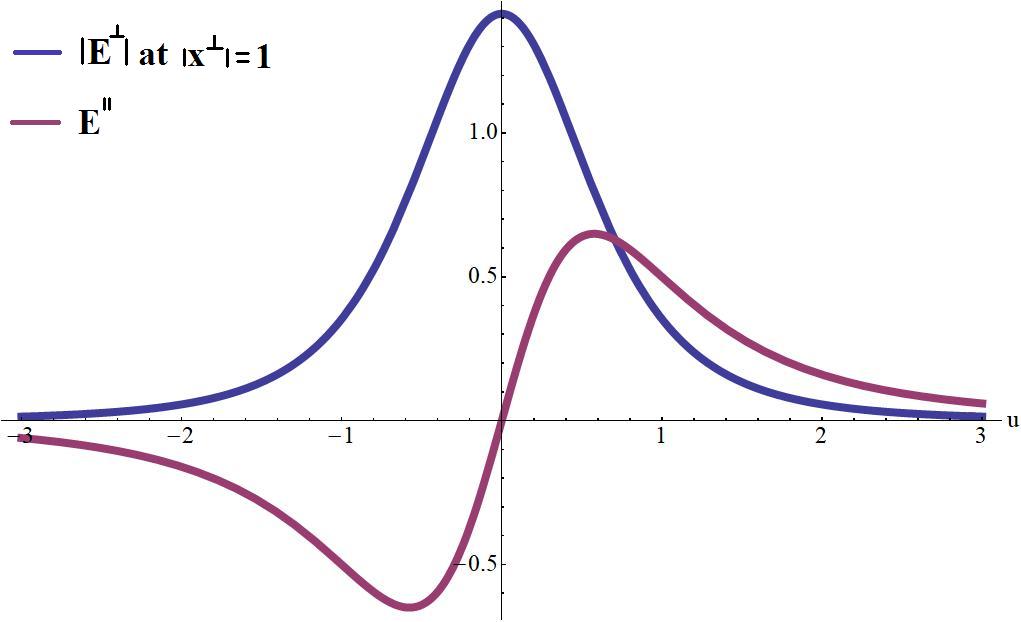} \hspace{1cm}
\includegraphics[width=0.45\columnwidth]{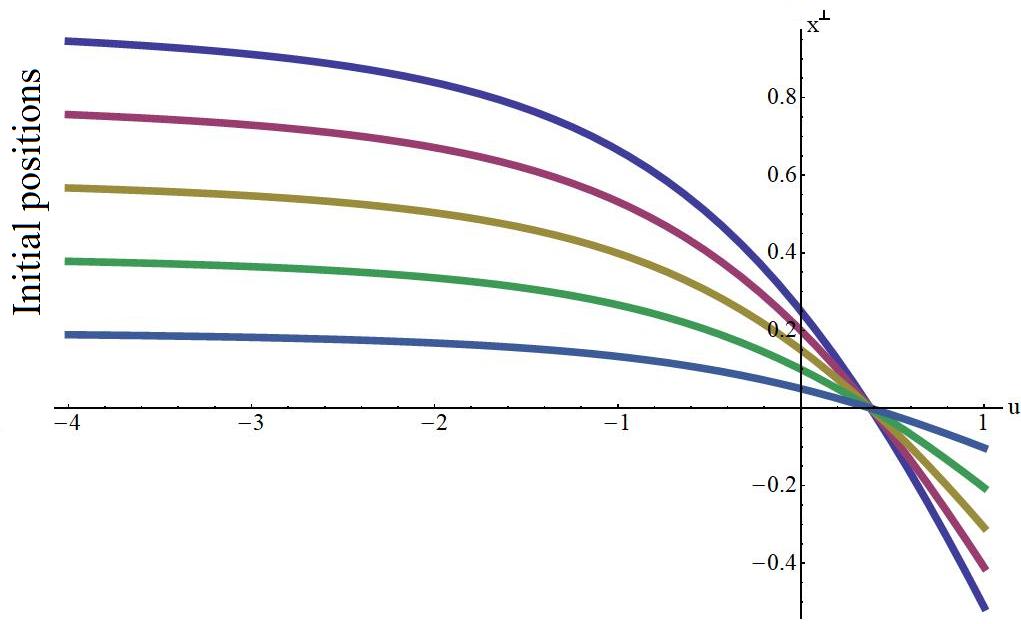}
\\
\end{center}
\caption{\small{The transverse and longitudinal electric field profiles  corresponding to   (\ref{e47}), i.e.  (\ref{e49}) with      $r=0$,  as well as  transverse part of  focusing solutions (\ref{e48}) ($e=-1$, $a=\epsilon=1$, $p_v=-1/3)$.}
\label{fig1}}
\end{figure}

Now, taking $H_{11}=H^{(1)}_{11}$ given by (\ref{e8})  one gets  
\be
\label{e46}
\mF(u)=\frac{-a}{u^2+\epsilon^2},
\ee  
and the  following electromagnetic vector field 
\be
\label{e47}
\vec E=\frac{\sqrt 2a	}{(u^2+\epsilon^2)^2}(x^1,x^2,\sqrt 2 u), \quad \vec B=\frac{\sqrt 2a	}{(u^2+\epsilon^2)^2}(-x^2,x^1,0),
\ee
 which was  studied, in another gauge,  in \cite{b10a}.  More precisely, it was shown there that   the transverse Lorentz equations are explicitly solvable; moreover,   there exist    solutions exhibiting the focusing property. Namely, for  the electromagnetic field (\ref{e47})   the  conditions (\ref{e21b})  lead to the following    solutions to  eqs. (\ref{e43a})  
  \be
  \label{e48}
 {\bf x}(u)=\frac{{\bf x}_{in}}{1-g}\left[1+\frac{ug}{\epsilon\sqrt{1-g}}\left(\frac\pi 2+\tan^{-1}(\frac{u}{\epsilon\sqrt{1-g}})\right) \right],
\ee
where 
\be 
\label{e48b}
g\equiv\frac{ ea}{-p_v\epsilon^2},
\ee
satisfies $g<1$. Now,  one can show that for $g<0$ there is a point $u_0$ such that $x^i(u_0)=0$, i.e.  focusing;   for $g>0$ there is no focusing. 
For example, taking $a=-\textrm{sgn}(e)$ one gets   the focusing case  $g=\frac{|e|}{\epsilon^2p_v}<0$. 
 The transverse and longitudinal electric field profiles of (\ref{e47}) as well as  solutions (\ref{e48}) are ploted in Fig. \ref{fig1}. 
\par  
For $g\geq 1$ the solutions are also   expressible in terms of  elementary functions (they consist of  composition of logarithmic and rational functions); however,   they exhibit  singularities  even though the starting potential is regular.   This fact  is related to the  coefficient  $eA_v-p_v$ appearing in front the second derivative in   eqs. (\ref{e43a})  and (\ref{e43b}). The condition  $g<1$  ensures that this coefficient is never zero.  Then  for continuous potential  the solutions of these equations  are regular. If  there are some points where this coefficient vanishes (here $g\geq 1$)  then  the order of the differential equations is at most one and then singularities can appear. This is the reason for which  starting from continuous potential we arrive, for some physical parameters,  at singular solutions.  Moreover,  the vanishing of this coefficient implies (see eq. (\ref{e42})) that reparametrization is valid only on some intervals (not globally  defined) which  makes considerations more  complicated.    In view of the above here and below  we skipped   such singular solutions as  physically  less transparent.
\par
In Ref.  \cite{b10a} it was suggested that  it would be interesting to analyse explicitly an example of the electric field  (\ref{e38}) with multiple field oscillations (in contrast to  (\ref{e47})). To this end  we consider a  counterpart of   the circularly polarized  plane gravitational  waves (\ref{e9}). Namely, let us  put  $\gamma=r\epsilon$, $r>0$ in $H^{(2)}$ (see (\ref{e9}) and (\ref{e9a})) and consider the profile 
\be
\label{e49}
H_{11}=H_{22}=\frac{a}{(u^2+\epsilon^2)^2}\cos(2r\tan^{-1}(u/\epsilon) ),\quad  H_ {12}=H_{21}=0.
\ee
Then the  function $\mF$ is  as follows:  for $r=1$
\be
\label{e50}
\mF(u)=\frac{au^2}{(u^2+\epsilon^2)^2},
\ee 
and  for $r\neq 1$
\be
\label{e51}
\mF(u)=a\frac{2\epsilon ru\sin(2r\tan^{-1}(u/\epsilon))+(\epsilon^2-u^2)\cos(2r\tan^{-1}(u/\epsilon))}{2\epsilon^2(r^2-1)(u^2+\epsilon^2)}+\frac{a\cos(\pi r)}{2\epsilon^2(r^2-1)}.
\ee
\par
Let us begin with  the case of   $r=1$.  Then 
\be 
\label{e51b}
\mE(u)=\sqrt{2}a\frac{\epsilon^2-u^2}{(u^2+\epsilon^2)^3},
\ee  
and  the transverse part of the  electromagnetic field is picked at $u=0$ and  has  two zeros 
{(in contrast to the profile discussed in   \cite{b10a}).  In  what follows we shall show that  the  trajectories  as well as focusing conditions can   be also  explicitly written   down. }
\par 
{To this end let us  consider } three cases distinguished by the value of the constant   $g$ given  by eq. (\ref{e48b}).
First, let  $g>0$. Imposing  (\ref{e21b}) one gets  the 	 solution 
 \be
 \label{e52}
{\bf x}(u)={\bf x}_{in}\left[1+\frac{gu}{\sqrt{4 g+ g^2}}\left(\frac{1}{\sqrt{A_-}}\left(\frac \pi 2+\tan^{-1}(\frac{u}{\sqrt{A_-}})\right)-\frac{1}{\sqrt{A_+}}\left(\frac \pi 2+\tan^{-1}(\frac{u}{\sqrt{A_+}})\right)\right)\right],
\ee 
where 
\be
\label{e53}
A_{\pm}\equiv \frac{\epsilon^2}{2}(2+g\pm\sqrt{4g+g^2})>0.
\ee
Using $g>0$ one can show,  after some calculations, that  there is no focusing.
\par
\begin{figure}[!ht]
\begin{center}
\includegraphics[width=0.45\columnwidth]{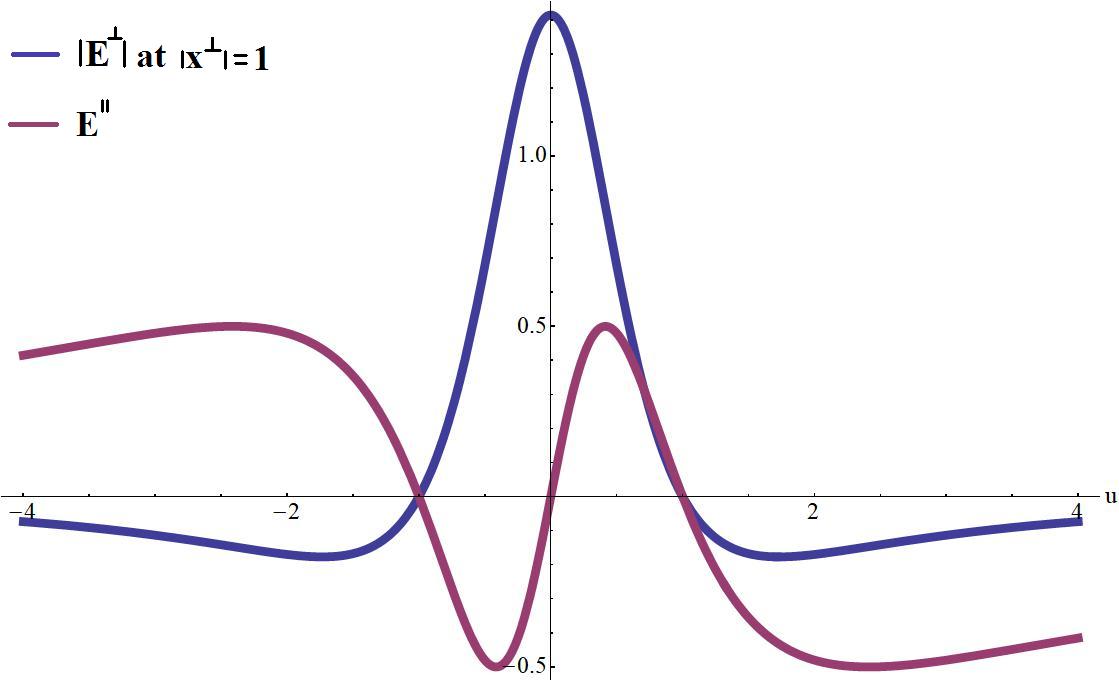}\hspace{1cm}
\includegraphics[width=0.45\columnwidth]{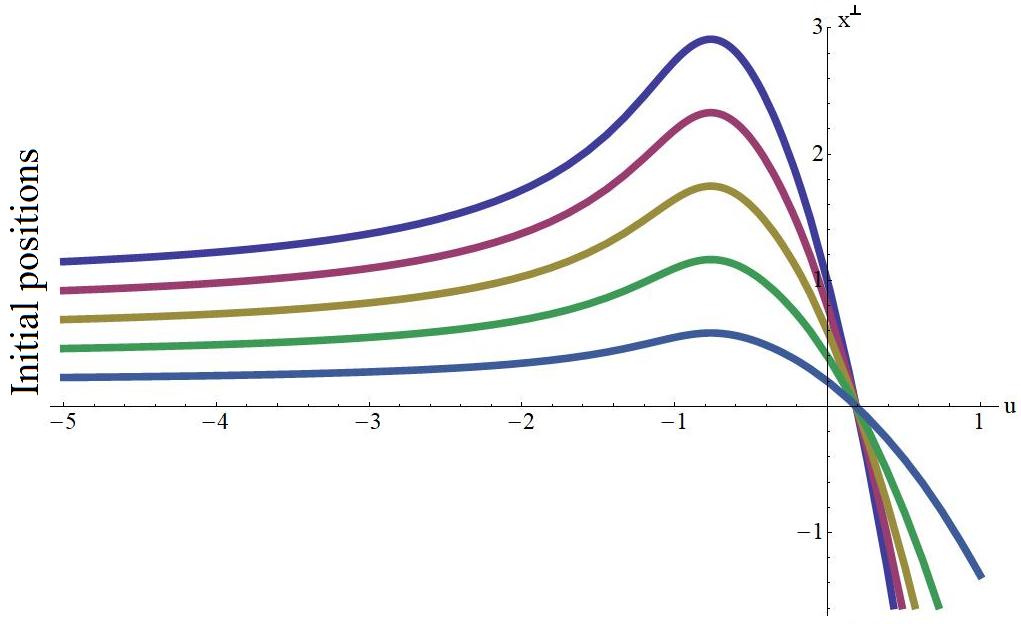}
\\
\end{center}
\caption{{\small The transverse and longitudinal electric field profiles  corresponding to  (\ref{e51b}), i.e.  (\ref{e49}) with      $r=1$,  as well as transverse part of  focusing solutions  (\ref{e54}) ($e=-1$, $a=\epsilon=1$, $p_v=-1/3)$.}
\label{fig2}}
\end{figure}

 
Next, when $0>g>-4$, we have 
\be
\label{e54}
\begin{split}
{\bf x}(u)={\bf x}_{in}&\left[1+\frac{ gu}{4 B}\left(\pi+\tan^{-1}\left(\frac{u+\epsilon\sqrt{-g/4}}{B}\right)+\tan^{-1}\left(\frac{u-\epsilon\sqrt{-g/4}}{B}\right)\right)\right.\\
&\left.-\frac{\sqrt{-g}u}{2\epsilon}\tanh^{-1}\left(\frac{\epsilon\sqrt{-g}u}{u^2+\epsilon^2}\right)\right],
\end{split}
\ee
where 
\be
\label{e55}
B=\epsilon \sqrt{1+g/4}.
\ee
Then there is a point $u_0$  where $x^i(u_0)=0$  for all initial points, i.e.  focusing.  
Let us note that taking $a=-\textrm{sgn}(e)$  one gets  $g=\frac{|e|}{p_v\epsilon^2}<0$;  however, only for some values of the particle parameters  the condition  $g>-4$ holds (in contrast to the electromagnetic field (\ref{e47}),  non-vanishing in the transverse direction). 
The transverse and longitudinal electric field profiles  corresponding to  (\ref{e51b})  as well as the   focusing solutions  (\ref{e54}) are illustrated in Fig. \ref{fig2}. 
\par
Finally, for $g\leq -4$ the solutions can  also  be expressed  in terms of elementary functions; however, they exhibit some singularities thus we skip them here (see the discussion after (\ref{e48b})). 
\par For  arbitrary function $\mF$     the integral in (\ref{e45})  cannot be explicitly computed; even when the functions $\mE $ and $\mF $ are both rational ones (e.g., $\mF$  given by (\ref{e49})  with the  non-negative integer $r$). Thus, in what follows we  give some sufficient conditions to ensure the focusing property of the electromagnetic field (\ref{e38}) and  next we apply them to the case (\ref{e49}).
\par 
To begin with, we should assume that the denominator in the integral (\ref{e45})  does not vanish anywhere
\be
\label{e55b}
e\mF(u)-p_v\neq 0;
\ee 
this can be achieved by a suitable choice of the  range of  parameters $e$ and/or  $p_v$ (since $\mF$ tends to zero at the null infinities). Next,	 we rewrite the general solution in the following form 
\be
\label{e56}
x^i(u)=C^i_1x_1^i(u)+C_2^i x_2^i(u)=C_1^iu+\frac{C_2^ip_v}{e\mF(0)-p_v} +C_2^iu\int_{-\infty}^u\frac{\mG(\bar u)d\bar u}{\bar u^2(e\mF(\bar u)-p_v)},
\ee    
where 
\be
\label{e57}
\mG(u)=\frac{-ep_v(\mF(0)-\mF(u))}{e\mF(0)-p_v}.
\ee
Then  the function under the integral (\ref{e56})   is well defined on the whole real line  since $\mG(0)=0$ and, by virtue of  (\ref{e44}),  $\dot \mG(0)=0$.  Now imposing the   conditions  (\ref{e21b}) one obtains $C_1^i=0$ and, by virtue of L'Hospital's rule,   the following  form of the solution
\be
\label{e58}
{\bf x}(u)={\bf x}_{in}\left(-u\int_{-\infty}^u\frac{\mG(\bar u)d\bar u}{\bar u^2(e\mF(\bar u)-p_v)}-\frac{p_v}{e\mF(0)-p_v}\right).
\ee   
Hence, the inequality
\be
\label{e59}
\frac{1}{(e\mF(0)-p_v)}\int_{-\infty}^\infty\frac{e(\mF(0)-\mF(u))du}{u^2(e\mF(u)-p_v)}>0,
\ee
together with (\ref{e55b})  imply a focusing point. For instance, the  condition (\ref{e59}) is satisfied    when   the function $e\mF(u)$ has the global maximum at $u=0$,
\be
\label{e60}
e\mF(0)\geq e\mF(u).
\ee 
\par 
One  can check that   in the case  of   the electromagnetic field  defined  by (\ref{e47})  and (\ref{e51b})  (equivalently,  (\ref{e49}) with  $r=0,1$) the criteria  (\ref{e55b}) and (\ref{e60})  yield the conditions  obtained  above.  Furthermore, applying these criteria to the  electromagnetic field  defined by  (\ref{e49}) with $0<r<1$ we arrive at the following  focusing  conditions 
\begin{align}
 \textrm{\quad for } \quad 0<r\leq \frac 12,&\qquad  g<0;\\
 \label{e61b}
\textrm{\quad for } \quad \frac 12<r<1, & \qquad \frac{2(1-r^2)}{\cos(\pi r)+r\sin(\frac{\pi}{2r})}<g<0.
\end{align}
Taking into account  the above  discussed   case $r=1$ we see explicitly  that  when the profile  of the electromagnetic field  vanishes at some points,   see   (\ref{e44b}) and (\ref{e49})  with  $\frac 12 <r\leq 1$, then there are   some additional  restrictions for the particle parameters (see (\ref{e61b})). Finally, it is worth to notice that (\ref{e61b}) gives   a new  Kober's-type inequality \cite{c1}.
\par
\begin{figure}[!ht]
\begin{center}
\includegraphics[width=0.45\columnwidth]{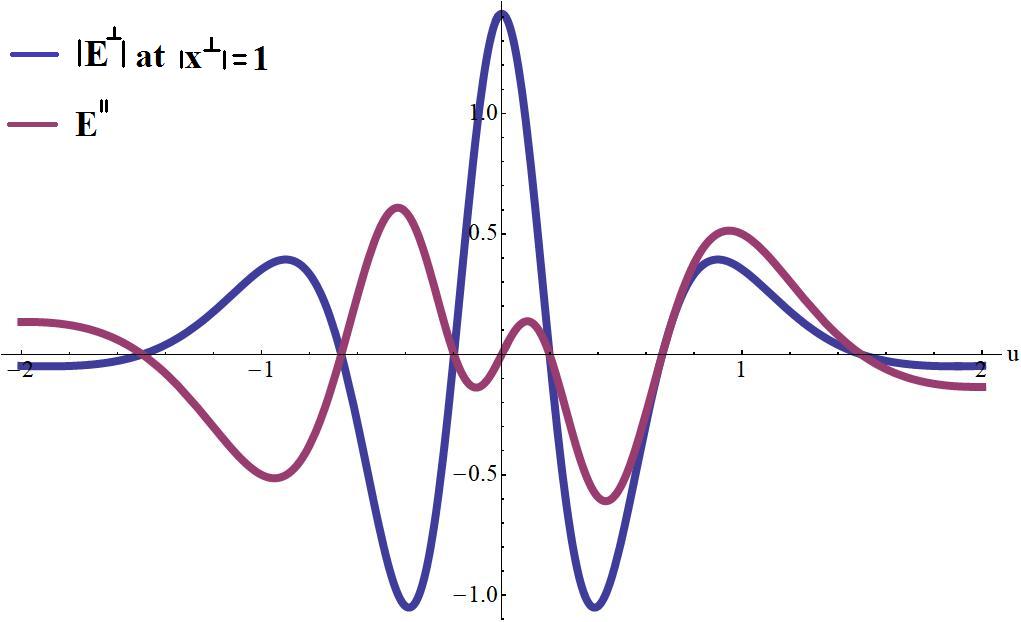}\hspace{1cm}
\includegraphics[width=0.45\columnwidth]{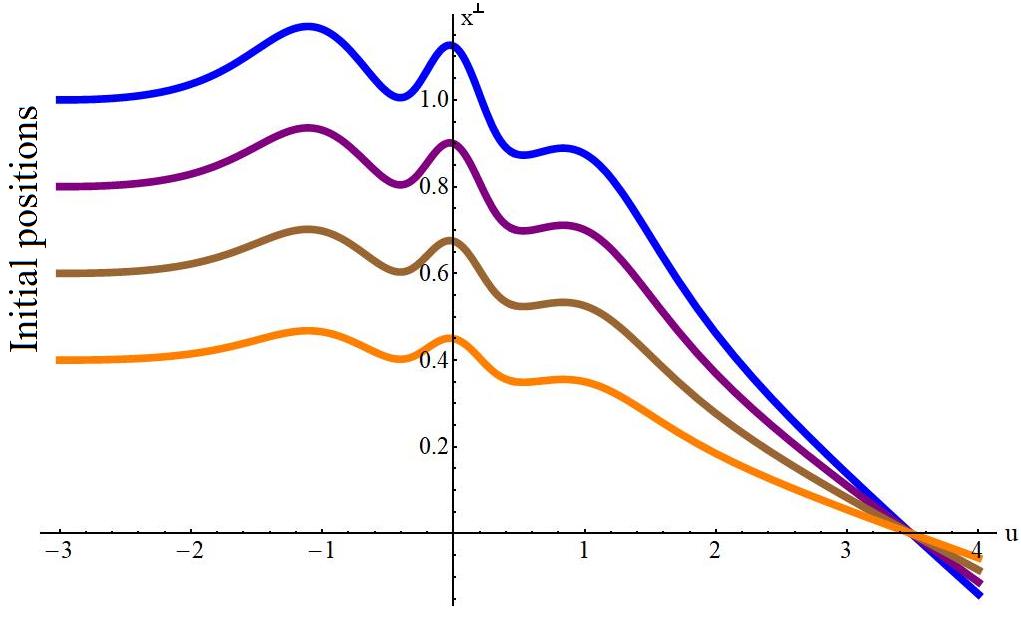}
\\
\end{center}
\caption{{\small The transverse and longitudinal electric field profiles  corresponding to   (\ref{e49}) with      $r=4$  as well as transverse part of  focusing solutions ($e=-1$, $a=\epsilon=1$, $p_v=-1/2)$.}
\label{fig3}}
\end{figure}

Finally, let us not that differentiating  (\ref{e58}) and using (\ref{e44})  as well as (\ref{e57}) one obtains the relation 
\be
\overset{..}{\bf x}=\frac{-p_ve\mE(u){\bf x}_{in }} {(e\mF-p_v)^2},
\ee
thus  the zeros of the electromagnetic profile $\mE$ are strictly related to the  inflection points of the solutions (\ref{e58}), see  Fig. \ref{fig1}-\ref{fig3} (e.g. for $r=0$  and $g<0$, Fig. \ref{fig1}, they are  concave).
 \section{Conformal symmetry and integrals of motion}
To obtain a  wider view on the results   discussed above   let us have a look on them from the integrals of motion point of view.  The discussion of the integrability  of the electromagnetic fields  of the type which appear in  previous sections     were partially elaborated  in Refs. \cite{b18,b10a,b10b}.  In particular, it was shown  there that such  systems  are superintegrable.  However, let us stress that   such   (super)integrability does not ensure  that the equations of motion are explicitly solvable. In view of this  the question is whether  there  are new integrals of motion (or even better --  symmetry generators)    which can help to find explicit solutions. Let us   study this problem in more detail for the electromagnetic vortices discussed in Sec. 3 and 4. To obtain the more complete picture of  this problem first, following Refs.  \cite{b18,b10a,b10b}, (however applying  our conventions of  the  signature  and gauge fixing)    we present  the  general discussion of the integrability. 
 \par
Let us start with the Lagrangian of a relativistic particle in  an electromagnetic background $A_\mu$   
 \be
 \label{r1}
 L=-m\sqrt{-\frac{d x^\mu}{d\tau}\frac{d x_\mu}{d\tau}}+eA_\mu\frac{d x^\mu }{d \tau}.
 \ee
 Since $L$ is homogeneous of the  first degree in velocities (it is invariant under  reparametrization)  the Hamiltonian (evolution generator)  of the particle vanishes. To remedy this problem  and   to  describe the  dynamical evolution one has to fix the gauge  by choosing  a time parameter $\tau=\tau(x^\mu)$ (of course, all choices  lead to equivalent results; however, some of them can be more useful for given  systems);  the most popular is the affine parametrization. 
On the other hand, for the discussed  systems the light-cone coordinates are more  suitable. In these coordinates  the Lagrangian and  the affinity  condition  take the form 
\be
L=-m\sqrt{-2\frac{d u}{d\tau}\frac{d v}{d\tau}-\left(\frac{d {\bf x}}{d \tau}\right)^2}+eA_u\frac{d u }{d \tau}+ eA_v\frac{d v }{d \tau}+e{\bf A}\frac{{d \bf x}}{d \tau}, 
\ee
\be
\label{r3}
2\frac{d u}{d\tau}\frac{d v}{d\tau}+\left(\frac{d {\bf x}}{d \tau}\right)^2=-1.
\ee
\par 
Now, let us assume that the  potential $A_\mu $ of the electromagnetic field  can be chosen as follows:   ${\bf A}={\bf A}(u,{\bf x})$,   $A_u=A_u(u,{\bf x})$  and  $A_v=A_v(u)$ (such a situation includes all above considered cases).  Then the  form of the potential    determines the function   $\tau(x^\mu)$ uniquely.   Indeed,   the canonical momentum $p_v$ conjugated to the $v$ coordinate is an integral  of motion. In consequence, the condition \eqref{r3} yields 
\be
\label{r4}
p_v=m\frac{du}{d\tau} +eA_v(u) =const,
\ee 
which  implies the form of the affine parameter $\tau=\tau(u)$. To make   this reparametrization invertible   one should assume that the derivative    $\frac{du}{d\tau}$ is of constant sign (positive  or negative);  Furthermore, for the free particle $p_v=p^u=(p^3-p^0)\sqrt 2<0$, thus  assuming that $A_v$   vanishes at  null infinities   one obtains $\frac{du}{d\tau}<0$. 
Then 
\be
\label{r5}
L=m\sqrt{-2\dot v -\dot {\bf x}^2}+eA_u+ eA_v\dot v+e{\bf A}\cdot \dot{\bf x}.
\ee       
Finally,  the condition \eqref{r3} takes the form 
\be
\label{r6}
\dot {\bf x}^2+\frac{m^2}{(p_v-eA_v)^2}=-2\dot v.
\ee
 Of course,  the above Lagrangian together with the condition \eqref{r6}  implies the suitable equations  of motion.
\par 
Now, we are in the  position to form the Hamiltonian formalism.  The  phase  space obtained is six-dimensional,  $(v,p_v)$ and  $({\bf x}, {\bf p})$ with the  canonical Poisson brackets, and the  Hamiltonian is  of the following form
 \be
 \label{r7}
 H=\frac{({\bf p}-e{\bf A})^2+m^2}{2(p_v-eA_v)}-eA_u.
 \ee
Moreover, let us note that combining eqs. \eqref{r4} and \eqref{r6} one obtains the relation $p_u=-H$.
\par 
Let us now  discuss the  problem of integrals of motion. First,  let us  consider the electromagnetic potential  corresponding to  the plane gravitational waves (\ref{e000}), i.e.  $A_u(u,{\bf x})=-{\bf  x}\cdot A(u) {\bf x},\quad {\bf A}=0,\quad A_v=0$, where $A(u)$ is a symmetric matrix.  Then the Hamiltonian \eqref{r7} yields the desired  equations of motion (\ref{e32a}) and (\ref{e32b}). As we indicated above $p_v$  is an  integral of motion.  However, it turns out that  \cite{b18}   the quantities
\be
 \label{r9}
I_{\bf k}= {\bf k}\cdot {\bf p}-p_v{\bf x}\cdot \dot {\bf k},
\ee
where ${\bf k}={\bf k}(u)$  is a  solution of the following set of eqs.  (cf.  eqs. (\ref{e32a})) 
\be
 \label{r10}
  \overset{..} {\bf k}=\frac{2e}{-p_v}A{\bf k},\\
\ee
are also integrals of motion.  Of course, $I_{\bf k}$ and $p_v$ are in involution. Moreover, by choosing four   independent solutions of the set of eqs. \eqref{r10}  and noting that the Wronskian of these solutions  is independent of $u$ (in consequence it can be determined  at one point) $I_{\bf k}$  yields four constants  of motion  forming two (independent) copies of the Heisenberg algebra. Summarizing, we have three integrals in involution and the   two additional integrals; thus the motion in the   these electromagnetic fields is maximally  superintegrable.  However,  let us stress that even such superintegrability does  not ensure the explicit solutions of the equations of motion (\ref{e32a}) and (\ref{e32b})  and does not select any special profile  $A$.
\par 
To better understand   the explicit solvability of the discussed electromagnetic vortices (\ref{e33}) and (\ref{e34})  let us recall  that with any Killing  vector field $K$, e.g.  a Poincare generator in the Minkowski spacetime,   one can associated  an integral of motion of the free particle.  Of course, when an  electromagnetic background is switched on then,  usually, this integral   is  not  a  conserved charge. However,  if the  Lie derivative of  the potential $\mA=A_\mu dx^\mu$ satisfies the following condition 
 \be
  \label{r11}
 \mL_K \mA=d\chi,
 \ee
 with   a function  $\chi$ (i.e.  $\mL_K  F=0$)  then 
 \be
  \label{r12}
 I_K=K_\mu P^\mu-e\chi=K_\mu(m \frac{d x^\mu}{d\tau}+eA_\mu)-e\chi,
 \ee
 is an integral  of motion.
\par  Now, let  us  assume that $K$ is a conformal vector field   with the conformal factor $2\psi$ and such that  \eqref{r11} holds. Then  along the  trajectories one obtains
 \be
  \label{r13}
 \frac{dI_K}{d\tau}= m\psi.
 \ee
 The key observation is that if  $\psi$  is a  function of $\tau$ then it  can be  rewritten as derivative of some function of $\tau$; in consequence one obtains a new, $\tau$ dependent,  integral of motion (for example, if $K$ is  a homothetic vector field satisfying \eqref{r11}  then $\psi=\psi_0$ is  a constant and consequently   $I_K -m\tau\psi_0$ is an integral   of motion). 
\par 
 Let us  apply the above  approach to  the electromagnetic  vortices discussed in Sec. 3 and 4.   To this end let us recall that the space of conformal vector fields on the  Minkowski spacetime is generated by the Killing
fields together with  three types (non-isometric) of conformal fields: radial,  special and general ones, see  \cite{b6b}.  In particular,  there is a standard special conformal vector $S$ (i.e. the one for which  the gradient of the conformal factor is a parallel null vector along $v$  coordinate); it is  of the  following form 
\be
S=u^2\partial_u-\frac1 2 {\bf x}^2\partial _v+u{\bf x }\cdot \nabla, 
\ee
and its  conformal factor  reads  $\psi=u$.  Since our aim is to  find  conformal vectors  satisfying conditions \eqref{r11}  for  the  potential  with $A^{(1)}$ (corresponding to (\ref{e8}) )   we add to $S$ some Killing vectors (here Poincare generators). Namely, by straightforward computations one checks that  the following conformal vector field (obtained by   adding  the $u$-translation generator)
\be
K^{(1)}=S+\epsilon^2\partial_u= (u^2+\epsilon^2)\partial_u-\frac1 2 {\bf x}^2\partial _v+u{\bf x }\cdot \nabla,
\ee 
satisfies  
\be
\mL_{K^{(1)}}\mA^{(1)}=0,
\ee
i.e.  eq. \eqref{r11}  with  $\chi=0$. 
This fact  perfectly agrees with  the gravitational  picture where $K^{(1)}$ is a conformal generator for the gravitational plane wave with the profile $H^{(1)}$, see \cite{b77}.  In consequence eqs. \eqref{r6} and \eqref{r13} lead to  the following integral of motion 
\be
I^{(1)}=-\frac{m^2\epsilon^2}{2p_v^2}- ea\frac{(x^1)^2-(x^2)^2}{p_v(u^2+\epsilon^2)}-(u^2+\epsilon^2)\frac{\dot {\bf x}^2}{2}-\frac1 2 {\bf x}^2+u{\bf x }\cdot \dot {\bf x},
\ee
which  is of the same  form as   the one  obtained   for the  plane  gravitational wave from  $K^{(1)}$, see \cite{b77}  (more precisely, they coincide after replacement $a\rightarrow \frac{2ae}{-p_v}$  in the gravitational case).  In summary, we see that   the conformal generator $K^{(1)}$ implies, in the case of  discussed electromagnetic vortices in the  Minkowski spacetime,  the same integral of motion (modulo  coupling constants) as $K^{(1)}$ for the  gravitational wave (\ref{e000}) with $H^{(1)}$  (but without the electromagnetic potential); the  term arising from the electromagnetic  potential coincides  with the additional term appearing in  $\dot v$ in the case of  gravitational waves.    This  immediately allows us to  understand  better the role of the conformal symmetry in the explicit solvability of  the equations  of motion for the electromagnetic vortices (\ref{e33}). Indeed, following Ref. \cite{b77}   $I^{(1)}$ can be rewritten as the sum of two independent Ermakov-Lewis invariants  \cite{a0a,a0b,a0c}, namely   
\be
{I^{(1)}}=-\frac{m^2\epsilon^2}{2p_v^2}-\frac{\epsilon}{2}\left[(\rho \dot x^1-\dot \rho x^1)^2+\frac{\Lambda_1 (x^1)^2}{\rho^2}\right]
-\frac{\epsilon}{2}\left[(\rho \dot x^2-\dot \rho x^2)^2+\frac{\Lambda_2 (x^2)^2}{\rho^2}\right],
\ee
where $\Lambda_i$ are defined by eqs.  (\ref{e11})  with replacement $a\rightarrow \frac{2ae}{-p_v}$ and $\rho$ is of  the form
\be
\label{r17}
\rho(u)=\frac{\sqrt{u^2+\epsilon^2}}{\sqrt{\epsilon}}.
\ee
Moreover, the function $\rho$ satisfies  the set of the  Ermakov-Milne-Pinney  equations  for the profiles $A^{(1)}$  and  $\tilde A^{(1)}$ (the latter one  is a diagonal matrix with the constant  elements  $-\Lambda_i$, $i=1,2$)  
  \be
  \label{r18}
 \overset{..}{\rho}I-A^{(1)}\rho=-\frac{\tilde  A^{(1)}}{\rho^3}.
 \ee
This information can be used to  find the   solutions of  the transverse  part  of the   equations of motion.   According to the  general  procedure, see  e.g. \cite{a5},  the transformation
\be
  \label{r14}
\frac{d \tilde u}{du}=\frac{1}{\rho^2(u)}, \quad {\bf x}=\sqrt{\epsilon}\rho(u)\tilde {\bf x},
\ee
should  relate the $u$-dependent linear oscillator,  defined by $A^{(1)}$, to the harmonic one with  the frequencies  $\Lambda_{1,2}$. In  our case, i.e. $\rho$   given by  (\ref{r17}),  the above formulae yield the Niederer transformation, eqs. (\ref{e12}),  and consequently the explicit integrability.   
Furthermore,  it turns out  (see \cite{a5}  and references therein) that    the Ermakov-Lewis  invariants can be interpreted as the ``classical"  energy in the new coordinates $\tilde {\bf x} ,\tilde u$. In our  case this leads to the identity 
\be
I^{(1)}= -\epsilon^2\left[\frac{m^2}{2p_v^2}+ E^{(1)}\right],
\ee
where 
\be
E^{(1)}=\frac12 \tilde {\bf x}'^2-\frac12\tilde {\bf x}\cdot\tilde H^{(1)} \tilde {\bf x}=\frac12 \tilde {\bf x}'^2+ \frac{\Lambda_1(\tilde x^1)^2}{2} + \frac{\Lambda_2(\tilde x^2)^2 }{2}.
\ee
and $\Lambda_i$  are defined by (\ref{e11})  with the mentioned above replacement.  In consequence, we obtain an  interpretation of the integral of motion associated with the proper conformal generator $K^{(1)}$.   
\par 
In the  case  of  the potential $A^{(2)}$ defined by the profile (\ref{e9})  we have to  add  to $K^{(1)}$ (and consequently to $S$) another  Poincare generator.  Namely,   direct computations show that the conformal vector  field 
\be 
K^{(2)}=K^{(1)}-{\gamma}(x^2\partial_1-x^1\partial_2),
\ee
satisfies 
\be
\mL_{K^{(2)}}\mA^{(2)}=0,
\ee
i.e.  eq. \eqref{r11}  with  $\chi=0$. Moreover, the conformal factor is the same as  for $K^{(1)}$, i.e. $\psi=u$.  In consequence, by virtue of eqs.  \eqref{r6} and \eqref{r13} one obtains the following integral of motion
\begin{align}
I^{(2)}&=-\frac{m^2\epsilon^2}{2p_v^2}- \frac{e}{p_v} (u^2+\epsilon^2){\bf x}\cdot A^{(2)}{\bf x}-(u^2+\epsilon^2)\frac{\dot {\bf x}^2}{2}-\frac1 2 {\bf x}^2+u\dot {\bf x }\cdot  {\bf x}-{\gamma} \dot {\bf x}\times {\bf x},
\end{align}
which again differs from the one obtained for the  plane gravitational   wave (\ref{e000})  with the profile (\ref{e9})  by the replacement  $a\rightarrow \frac{2ae}{-p_v}$ (cf.  results in \cite{b77}).  Furthermore, the function $\rho$ and  profile $A^{(2)}$ satisfy  the    Ermakov-Milne-Pinney  type equation  (similar to  eq. (\ref{r18})). In consequence, they lead to   the transformation  (\ref{e2}), i.e. allows  to find explicit solution of the transversal equation (\ref{e32a}). 
Finally, following \cite{b77}   the integral of motion $I^{(2)}$   can be interpreted as the sum  total energy for the system defined by the equations of motion (\ref{e17}), i.e. 
\be
I^{(2)}=-\epsilon^2\left[\frac{m^2}{2p_v^2}+ E^{(2)}\right],
\ee
where 
\be
E^{(2)}= \frac12 ({\bf y}')^2+\frac 12\Omega_+(y^1)^2+\frac 12 \Omega_- (y^2)^2,
\ee
and  $\Omega_\pm$ are given by eqs. (\ref{e18})  with  $a\rightarrow \frac{2ae}{-p_v}$.
\par 
Summarizing, we  have shown that   some  special conformal transformations of the Minkowski spacetime  can be chosen in  such a way that they  preserve  electromagnetic fields and consequently generate  new integrals of motion.  In the presented cases these integrals  can be interpreted in terms of  the  Ermakov-Lewis invariants  (Ermakov-Milne-Pinney equations) leading to   the  Niederer transformation and, consequently,  explicit solutions in terms of new variables. Moreover these integrals can be interpreted as  total energy of the  dynamical system in   new  coordinates.  All these observations coincide with the gravitational  picture where  these conformal vector fields survive  for the plane gravitational waves (\cite{b77} and references therein)  obtained by means of Kerr-Schild ansatz; the difference is that  instead of the electromagnetic potential  there is suitable term in the gravitational metrics.   
\par
Finally,  since the Kerr-Schild ansatz can be extended to  the fixed backgrounds (see  Sec. 3) this suggests that  the  above   observations concerning the conformal vector fields $K^{(1,2)}$  in the Minkowski spacetime  should be extended to the  electromagnetic potential $A^{(1,2)}$ in the fixed   $g^{(1,2)}$ backgrounds. This  fact can be checked directly  by   noting that the   condition \eqref{r11}  is metric independent. This coincides with the results obtained in Sec. 4.
\par
Above we discussed the integrability  problem for  the electromagnetic field  defined by the potential (\ref{e23}).   Finally,  let us  review  these issues for  the  electromagnetic field considered in Sec. 5, i.e. defined by 	
\be
\label{r15}
A_u(u,{\bf x})=-H_{11}(u){\bf  x}^2,\quad {\bf A}=0,\quad A_v(u)=2\int_{-\infty}^u\overline uH_{11}(\overline u) d\overline u,
\ee
where $H_{11}$ is an arbitrary function such that the last integral exists.
Then the Hamiltonian  formalism  is given by   \eqref{r7} together with   the above mentioned Poisson brackets (leading directly to  eqs. (\ref{e43a}) and (\ref{e43b})).  Also in this case  the canonical momentum   $p_v$ is a  constant  of motion; the   problem of  other integrals of motion of such system was discussed (in slightly  different gauge) in Refs. \cite{b10a,b10b}. Namely, it was shown that each of  the   two  following Poincare generators $K^{(i)}$ 
\be
\label{r16}
 K^{(i)}=u\partial_{i}-{ x^i}\partial _v,\quad i=1,2;
\ee
satisfy the condition \eqref{r11}  with the functions $\chi^{(i)}$  of the  form
\be
{\chi^{(i)}}=-A_v(u)x^i, \quad i=1,2.
\ee
 In consequence, the  Killing vectors \eqref{r16}  generate two, $u$-dependent,  integrals of motions 
\be
I^{(i)}=-x^ip_v+up_i+ex^iA_v(u), \quad i=1,2.
\ee
These integrals together with the previous one, $p_v$,  are in involution, thus  the system governed by the potential \eqref{r15} is integrable. Furthermore, it was  shown that there is an additional integral of motion (which is not a polynomial in $p_v$) what makes  this system (minimally) superintegrable.  However, as in the case of the  vortices   discussed  above   this superintegrability does not ensure the explicit form of solutions.  So the question is whether  one  can find other symmetry generators which distinguish the fields  discussed in Sec. 5 and  to  give more transparent explanations of the explicit solvability  or, in general,  a more detailed  discussion of the role of integrals of motion associated with the conformal symmetry (see the recent Refs. \cite{d1,d2}). These  issues   are left  for further investigations.
\section{Conclusions and  outlook}
\label{s7}
Let us summarize. In the present work, using the idea of the classical double copy, we  construct null electromagnetic fields  which are explicitly solvable and  directly generalize the electromagnetic vortices considered, in the context of singular optics,   in Ref. \cite{b9}
{(in contrast to the latter ones, they form some pulses and  the Dirac delta limit of the profile can be easily done). 
} Since these results are  strictly related to  the notion of  Niederer's transformation we analyse  the geometric  extension of the latter  in the case of the plane gravitational spacetimes endowed with  the crossed electromagnetic  fields (\ref{e25}) and ({\ref{e26});  in this approach the transverse part of the Lorentz  force equations transformed by the Niederer map can be obtained by means of a new metric (conformally related to the  initial one)  and a new, crossed,   electromagnetic field.  In consequence, for   some special  cases, related to conformal symmetry of spacetimes,  the transverse Lorentz equations  on a curved spacetime can be also analytically solvable. 
We also showed that these results possess their origin in additional integrals of motion  associated with  conformal generators  preserving the discussed vortices and plane gravitational waves.
\par 
In the second part  we showed that  the electromagnetic backgrounds which capture some essential features of the transverse magnetic laser  beam   near the beam axis, proposed in Ref. \cite{b10a}, can also emerge from some gravitational counterparts. Moreover, one of the latter ones  leads to the   electromagnetic profile  with zeros  and  is analytically solvable in the transverse directions;
such a situation allowed us to  discuss explicitly the focusing conditions. 
 Since the focusing properties of electromagnetic fields seem to be especially interesting; thus we  gave  some criteria and  apply them to the  electromagnetic counterparts of  some gravitational metrics. 
Finally, let us note that the form of the electromagnetic fields discussed   does not   suggest the  solvability of  the corresponding Lorentz force equations; however, 
in view of the   role of  conformal symmetry in gravity  and the double copy conjecture the integrability of dynamics governed by these special electromagnetic fields arises naturally (the existence of explicit solutions to the geodesic equations coincide with the existence of the explicit solutions to the Lorentz force equations). Such  a situation  suggests deeper relations between both gravitational and gauge theories in the double copy approach.    
 \par
 The results   obtained can be extended in various directions. As usual the   exact solvability at  the classical level should  have  its  reflection at  the quantum level. Thus   it would be interesting  to consider the quantum picture of the results obtained  here, including both gravitational and electromagnetic backgrounds (see e.g. \cite{b10a,b12b}). This is especially interesting in view  of the double copy and recent results  presented   in Ref. \cite{b19} for gravitational sandwiches.  Furthermore, let us recall that  the Penrose limit of spacetimes yields the plane gravitational waves;   thus, the question is which of ones correspond to the  distinguished plane wave spacetimes $g^{(1,2)}$ (cf. \cite{b4b,b6f,b6ff}).   Moreover, the  analytical  solutions obtained   can be also  useful in the recent studies concerning some aspects of  optical effects in  the nonlinear  plane gravitational  waves \cite{b22a,b22b} as well as trapping problems in gravity \cite{b18,b18b,b18c}.  Finally, following  Refs. \cite{b11a,b11b} and \cite{b23a}-\cite{b23b}, we hope that  they can be also useful in  the study of the light-matter interaction, especially for  strong focusing of short or intense laser pulses.
\vspace{0.5cm}
\par
{\bf Acknowledgments}
\par 
The  authors would like to thank    Piotr  Kosi\'nski for helpful  discussions as well as Peter  Horvathy  for useful remarks and suggestions.   Comments of Cezary Gonera, Joanna Gonera and Pawe\l   \ Ma\'slanka  are  also  acknowledged. 
 We are also  grateful to the anonymous referees  for  a number of useful suggestions which, we feel, improved substantially the readability of the paper  as well as inspired some  new results.
This   work  has   been partially   supported   by   the   grant  2016/23/B/ST2/00727  of  National  Science  Centre, Poland. 
 
\end{document}